\nonstopmode

\documentstyle[12pt]{article}

\input{psfig}

\def\FIGREF#1{\ref{#1}}

\newenvironment{equl}[1]{\begin{equation}\label{#1}}{\end{equation}}
\newenvironment{eqular}[1]{\begin{eqnarray}\label{#1}}{\end{eqnarray}}
\def\CITE#1{{\cite{#1}}}

\def\mxnote#1{}
\def\mxnote#1{\marginpar{\tiny ~~#1 }}
\def\REF#1{(\ref{#1})}
\def\xnote#1{}
\def\braket#1{\left|#1\right>}
\newcommand{\kk}{{\vec{k}}}
\newcommand{\qq}{{\vec{q}}}
\newcommand{\nn}{{\vec{n}}}

\def\srv#1{\langle #1 \rangle}

\def\sT0#1{\langle T_\tau [#1] \rangle_0}
\def\kr{^\dagger}
\def\ide{\rightarrow}
\def\spup{\uparrow}
\def\spdn{\downarrow}
\def\veps{\varepsilon}

\def\FAZA#1#2{{ie\over \hbar c}\int_{#1}^{#2}{{\vec A} \cdot {\vec dl} } }

\begin{document}

\begin{center}
\Large
% Raman spectrum and charge fluctuations \\
% in the copper-oxide superconductors
 The effect of large $U_d$ on the Raman spectrum
 in the copper-oxide superconductors
\end{center}

\vskip 0.5cm

\leftline{H.~Nik\v si\' c$^{\rm a}$, E.~Tuti\v s$^{\rm b}$
	and S.~Bari\v si\' c$^{\rm a}$
}

\vskip 0.5cm

\leftline{$^{\rm a}$Department of Physics, Faculty of Science,
	Bijeni\v cka 32, POB 162, Zagreb, Croatia
}

\leftline{$^{\rm b}$Institute of Physics,
	Bijeni\v cka 46, POB 304, Zagreb, Croatia
}

\vskip 0.5cm

{
\centerline{ABSTRACT}
\vskip 0.5cm

The effect of the charge fluctuations on
the electronic and Raman
spectrum  of high temperature
superconductors is examined, using the
slave boson approach to the large
Coulomb repulsion $U_d$ on the copper site.
Instead of the saddle
point approximation $\srv{b}\ne 0$ for the slave boson,
characteristic for various $N\ide\infty$
approaches, we confine ourselves to
the non-crossing
approximation  (NCA) diagrams for the Green functions.
	In this way
the effects
of charge fluctuations
and of the constraint of no double occupancy on the copper site
on the shape of
the electronic spectrum are studied primarily,
while the slave boson
diagrams responsible for
copper-copper spin correlations are
intentionally not included.
%4
{  The novel feature of the charge fluctuation dynamics
shows up in the slave boson  spectrum as the plateau extending
over the range of 1eV below
$\omega=0$. This is further reflected in
the electronic Raman spectrum that we calculate
for various combinations of the
polarization of the incoming and
scattered light.
The Raman spectrum shows the characteristic featureless behaviour up to
frequencies of the order of
1eV while the polarization dependence is also in
qualitative agreement with experiment.
}

\vskip 0.5cm

{}~~~~RUNNING TITLE~:~The effect of large $U_d$ on the Raman spectrum

{}~~~~KEYWORDS~:~Raman scattering, electronic structure, normal state
properties

\vfill
\eject

\section{Introduction}

The attention which the copper-oxide
superconductors have been attracting in recent
years is due to their
extraordinarily high  transition temperature to
the superconducting state.
	The interest in normal state
properties stems from the belief that a
better understanding of these
properties could give a clue to the superconducting
mechanism itself. The
Raman spectrum, and especially its continuous,
almost featureless, background spreading up to
the frequencies of the order of 1eV, is one of
the most remarkable of
these properties \CITE{SHM1,SHM2}.
	Considering the
energy scales involved it is natural to ascribe
this background to the electronic excitations in
the system.
	In this work we propose a microscopic theory
of this phenomenon, which is
believed to be a common feature of all
copper-oxide superconductors.

%\mxnote{ET rewrite: \\ NFS-krit \\ mag posebno}
	Several possibilities for the origin
of the Raman background were discussed
previously in the literature.
	We comment briefly on some
characteristic ones.
First, there are theories in
which the Raman background is attributed
to particular band crossing
effects \CITE{ZawadowskiPRL}.
	However, the one-dimensional bands used in the particular
case of YBa$_2$Cu$_3$O$_{7-y}$ (where they come from
the CuO chains) are not expected to be a
feature of the copper-oxide superconductors in general.
%(e.g. not in La$_{2-x}$M$_x$CuO$_4$)
	Second, the ``Nested Fermi Surface'' (NFS)
approach \CITE{VirostekRuvalds}
starts from the
assumption that parts of the Fermi surface
have a pronounced nesting property
and proceeds by considering the effect of a
weak Coulomb interaction.
	While this approach is certainly
an interesting one, various Fermi surface experiments and band
calculations indicate that the Fermi surface geometry
varies significantly among the HTSC materials,
which show the featureless Raman background.
	Third, the possibility was also considered
that the magnetic excitations
(paramagnons, related to the antiferromagnetic phase
of the pure, undoped compound and common to most HTSC
materials) are related to the Raman background,
 for instance through the coupling of light to the two-magnon
excitations.
	However, the magnetic
exchange constant $J$, measured by the polarized
neutron scattering in pure materials,
is not big enough (e.g.
$J\approx 1700K$ in  YBa$_2$Cu$_3$O$_6$,
\CITE{RossatMignot} and $J\approx 1400K$ in
La$_2$CuO$_4$ \CITE{Keimer92}) to account
for the excitations of the order of
$8000cm^{-1}$.
	Furthermore, the magnon (paramagnon) excitations
presumably become more damped as the doping level is increased
and the effective $J$ is significantly suppressed \CITE{RossatMignot},
which has been considered and qualitatively
understood by several theoretical approaches.
	On the other hand, it seems that the magnetic
excitations indeed contribute to the Raman background
around $4000cm^{-1}$.
	This contribution is pronounced
\CITE{HN1,HN2,HN3,HN4}
in pure and slightly doped materials and fades
out on further doping.
	In fact it seems that
this contribution is {\it superimposed}
on an almost uniform, featureless background extending
up to $\omega\sim$~1eV, present in superconductors.

	Therefore, in order to understand
the origin of this background,
we turn towards the role of charge
fluctuations and the effect of a strong Coulomb interaction,
present in most models of HTSC's.
	The possibility that the charge fluctuations are
responsible for the special properties  of the
normal phase was suggested by several authors
from the very beginning.
	Varma and collaborators
were seeking for the microscopic origin of their
marginal Fermi  liquid phenomenology in this
domain.
	However, their approach was essentially
a weak coupling one while most of the model
parameter evaluations give an
on-site repulsion $U_d$ on the copper atom
 which is large relative to the
copper-oxygen hybridization,
$U_d\gg t_0$.
  Therefore, we use a strong coupling approach
to calculate the effect of  large $U_d$ on the Raman spectra.
  However, the interatomic Coulomb
repulsion term $V_{pd}$, important in their
theory, is not included in the present calculation.
	The effect of the local constraint
imposed by the Coulomb repulsion
on the Raman spectrum is essential in our approach.
	In that respect it shares some major
physical points with the recent work by Bang \CITE{Bang}
in which the {\it t-J} model is considered
within the more usual approach of the
saddle point approximation plus some fluctuation corrections
in the slave bosons.

\section{Model and method}

   Following many previous works on the copper-oxide
superconductors
(refs. \CITE{KLR88,KLA89,NewnsCo} and  many others)
we treat the large Coulomb repulsion $U_d$ in
the  $U_d\ide\infty$ limit using the slave boson
formalism \CITE{Barnes,Coleman84,ReadNewns}.
  The application of this formalism is usually
accompanied by the immediate use of the saddle
point approximation $\srv{b}\ne 0$ for the
slave boson field $b$.
  This approach has successfully accounted for some
electronic properties of the system such as the
in-gap ``band'' containing the Fermi level, the
large electronic mass and the electronic
dispersion of the form found in  ARPES
experiments.
   Furthermore, the integration of the
high energy (few eV)  fluctuations
($\omega\sim\veps_p-\veps_d$)
of the boson field around the  saddle point adds
one more important ingredient to the physics of
narrow band fermions - the copper-copper
magnetic interaction $J_{dd}$ \CITE{SiLevin}.
   However, it is the intermediate
frequency range, $\omega<$~1eV, that  must
be primarily considered for the explanation of the
Raman  background.
   In this region of frequencies, the
saddle point approximation does not seem to be an
advantageous starting point.
   In particular, as with any
system of noninteracting fermions,
the system of effective fermions
that emerges on the saddle point level
has no intraband $\kk=0$ excitations, so that
self-energy corrections for fermion lines and vertex
corrections in the Raman scattering diagrams
are required in order to get
a finite Raman response for the frequencies of interest.
{ Furthermore, the spectrum of the
Gaussian fluctuations  of the slave
boson field around their saddle point
value shows \CITE{bMFAsp}  no sign of the continuum
present in the Raman scattering.
}
Also, some fundamental
difficulties of the saddle point
approach \CITE{MFAprobl} are accompanied by
some more practical ones.
    For example, the $\kk$-dependence of
the boson  Green function  and the
corresponding $\kk$-space integrals considerably
complicate the calculation beyond the
saddle point approximation
{ requiring extensive numerical calculations. }
Actually, the intersite correlation
of the bose field  should  be absent due to
the local gauge invariance of the model
in the slave boson representation, and to
the slow fluctuations of the boson phase \CITE{Coleman87}.
	Therefore, we do not use the saddle point
approximation for the boson field $b$
in our calculations.
	  Instead, we consider the high order diagrams in
the boson field, in which the electronic density
of states, similar in shape to
the one obtained  through the
 saddle point approximation (with the
aforementioned in-gap resonance), and the
electronic damping, emerge  on  equal footing
\CITE{msimple}.
  This approach is a direct extension
of the treatment of Kroha et al. \CITE{Kroha92}
of the Anderson impurity problem.
 	The model that we use for holes in $CuO_2$
planes  is the well known {\it p-d} model
\CITE{Emery87},
\begin{eqular}{Hsve}
H & =  & H_0 + H_{hyb}  + H_d ~, \\
H_0 & = & \sum_{\nn,\sigma}
\{\epsilon_p (c_{x\nn\sigma}\kr c_{x\nn\sigma}
+ c_{y\nn\sigma}\kr c_{y\nn\sigma} )
+ \epsilon_d d_{\nn\sigma}\kr  d_{\nn\sigma}\}  ~, \\
H_{hyb} & = &  \sum_{\nn,\sigma} t_0[d_{\nn\sigma}\kr
(c_{x\nn\sigma}+c_{y\nn\sigma}
-c_{x\nn-1_x\sigma}-c_{y\nn-1_y\sigma}) + h.c.]  ~, \\
\label{H_hyb_cd}
H_d & = &\sum_\nn
 U_d { d_{\nn \spup  }\kr d_{\nn \spup  }
 d_{\nn \spdn}\kr d_{\nn \spdn} }  ~,
\end{eqular}
	in which,
apart from the Coulomb repulsion $U_d$,
the parameters $\epsilon_d$, $\epsilon_p$
and $t_0$ denote respectively
the energies of the hole in  Cu $3d_{x^2-y^2}$
and O $2p(\sigma)$ orbitals
and the copper-oxygen hybridization.
	Transformed to the $\kk$-space, $H_{hyb}$,
acquires the following form,
with the lattice constant set to unity:
\begin{eqular}{Hhyb}
H_{hyb} & = & \sum_{\kk,\sigma} \{
d_{\kk\sigma}\kr [c_{x\kk\sigma} t_{x\kk}+c_{y\kk\sigma} t_{y\kk}]
+[c_{x\kk\sigma}\kr t_{x\kk}^*+c_{y\kk\sigma}\kr t_{y\kk}^*]d_{\kk\sigma} \} ~,
\\
&   & t_{x\kk}\equiv t_0 (1-e^{-ik_x}) \mbox{~,~~~~}
t_{y\kk}\equiv t_0 (1-e^{-ik_y})
\mbox{~.~~~~}
\end{eqular}
	The Hamiltonian simplifies on introducing
the bonding and non-bonding
combinations of the hole operators on oxygen sites,
\begin{equl}{bonabon}
p_\kk\equiv c_{b\kk}\equiv{t_{x\kk}c_{x\kk}+t_{y\kk}c_{y\kk}
 \over \sqrt{\mid t_{x\kk}\mid^2+\mid t_{y\kk}\mid^2} }
\mbox{~,~~~~}
c_{n\kk}\equiv{t_{y\kk}^*c_{x\kk}-t_{x\kk}^*c_{y\kk}
 \over \sqrt{\mid t_{x\kk}\mid^2+\mid t_{y\kk}\mid^2} }
\mbox{~.~~~~}
\end{equl}
Only the bonding combination  has an overlap
with the copper orbitals. In terms of these
operators, the part of the Hamiltonian which
describes noninteracting holes becomes
\begin{equation}
H_0  = \sum_{\kk,\sigma} \left[  \veps_p\left ( p\kr_{\kk,\sigma}p_{\kk,\sigma}
+
c\kr_{n\kk,\sigma}c_{n\kk,\sigma} \right) +  \veps_d
d\kr_{\kk,\sigma}d_{\kk,\sigma}
  \right]
\mbox{~,~~~~}
\end{equation}
\begin{equation}
H_{hyb}  =  \sum_{\kk,\sigma}  t(\kk)[d_{\kk\sigma}\kr p_{\kk\sigma}
+p_{\kk\sigma}\kr d_{\kk\sigma} ] \mbox{~,~~~}
t(\kk)\equiv\sqrt{\mid t_{x\kk}\mid^2+\mid t_{y\kk}\mid^2}
\mbox{~.~~~~}
\end{equation}
Therefore, from now on we omit the non-bonding states from
our discussion \CITE{nbRaman}.

	 In the slave boson representation \CITE{Barnes} in
the $U_d\ide\infty$ limit, the fermion operator $f\kr_{\nn\sigma}$
is introduced to describe a singly occupied, and
the boson operator $b\kr_\nn$ is introduced
to describe  an empty orbital state
at the copper site $\nn$.
 	 The substitution $d_{\nn\sigma}\ide b\kr_\nn
f_{\nn\sigma}$, which leads to the Hamiltonian in the
new representation, is accompanied  by a
constraint, which separates the physical states
from many non-physical ones in the enlarged
Hilbert space.
	In terms of the number operator $Q_\nn$,
\begin{equation}
Q_\nn\equiv\sum_{\sigma} f_{\nn\sigma}\kr f_{\nn\sigma}
+b_\nn\kr b_\nn~,~~~
\end{equation}
which commutes with the Hamiltonian, this
constraint is written as
\begin{equl}{Qn1}
Q_\nn=1~.~~~
\end{equl}
In the path integral formulation, the constraint is
introduced  through the
``Lagrange multiplier'' $\lambda_\nn$
in the partition function \CITE{Coleman87}
\begin{equl}{ZQ}
Z(Q_\nn=1)=\left\{ \prod_\nn
\int_{-i\pi/\beta}^{i\pi/\beta}d\lambda_\nn\right\}
        Z(\{\lambda\}) e^{ \beta \sum_\nn 1 \cdot \lambda_\nn}
\mbox{~,~~~~}
\end{equl}
\begin{equation}
Z(\{\lambda\})\equiv
  \int{{\cal D}b\kr {\cal D}b{\cal D}f\kr {\cal D}f{\cal D}p\kr {\cal D}p}~
        e^{-\int_0^{\beta}{\cal L}(\tau) d\tau}
\mbox{~,~~~~}
\end{equation}
\begin{eqular}{lagrange}
{\cal L}(\tau) & \equiv &  \sum_{\nn,\sigma}
{f_{\nn\sigma}\kr ({\partial \over \partial \tau}+\epsilon_d)f_{\nn\sigma}}
                     +\sum_{\nn,\sigma}
{p_{\nn\sigma}\kr ({\partial \over \partial \tau}+\epsilon_p)
p_{\nn\sigma}}
+\sum_{\nn}{b_\nn\kr {\partial \over \partial \tau}b_\nn} \\
               &   & +   H_{hyb} +\sum_{\nn}\lambda_\nn
(\sum_{\sigma}{f_{\nn\sigma}\kr f_{\nn\sigma}+b_\nn\kr b_\nn)} \nonumber
\mbox{~.~~~~}
\end{eqular}
The saddle point approximation for the
integration over $\lambda_\nn$ in \REF{ZQ}
fixes $\lambda_\nn$ to the value $\Lambda$ which is, due to
the translational  symmetry of the system, equal
on all copper  sites. Within this
approximation, the constraint \REF{Qn1}
is satisfied {\it on the average},
%
%\begin{equation}
% \srv{Q_\nn}_{(\lambda_\nn=\Lambda)}=-{\partial \over \partial \lambda_\nn}
%\ln(Z(\{\lambda\})) \mid_{(\lambda_\nn=\Lambda)}=1~.~~~
%\end{equation}
%
\begin{equation}
 \srv{Q_\nn}=-{1\over \beta}{\partial \over \partial \lambda_\nn}
\ln(Z(\{\lambda\})) \mid_{(\lambda_\nn=\Lambda)}=1~.~~~
\end{equation}
Following the approach which was previously tested
by Kroha et al. \CITE{Kroha92} on the $U\ide\infty$
Anderson model, with  the result that it is in
good agreement with those obtained by
other methods, we neglect the
fluctuations of $\lambda_\nn$ around $\Lambda$.
This means that  the states with $Q_\nn$ different
from unity become allowed.
However, the effective Hamiltonian,
\begin{equl}{HL}
\begin{array}{ccc}
 & H & =  \displaystyle
\sum_\kk \{\sum_\sigma [\epsilon_p p_{\kk\sigma}\kr p_{\kk\sigma}
  +(\epsilon_d+\Lambda)f_{\kk\sigma}\kr f_{\kk\sigma}]
  +\Lambda b_\kk\kr b_\kk \mbox{~~~~~~~~~}
\\ [3mm]   \displaystyle
 &   & \displaystyle +{t(\kk)\over \sqrt{N}} \sum_\sigma [ p_{\kk\sigma}\kr
 \sum_{\qq}b_{\qq}\kr f_{\kk+\qq\sigma}
+ \sum_{\qq} f_{\kk+\qq\sigma}\kr b_{\qq}p_{\kk\sigma} ] \}
\mbox{~,}
\end{array}
\end{equl}
with $\lambda_\nn$ fixed
to $\Lambda$, conserves $Q_\nn$, so the states
with different $Q_\nn$ are not dynamically mixed
($N$ stays for the number of unit cells in the crystal).
	  Therefore, we expect that fixing
$\srv{Q_\nn}=1$ in our calculation (which itself
conserves the $Q$-charge)  will predominantly
extract the dynamics of the system with $Q_\nn=1$.

	It can be easily checked
%\CITE{AppA}
that, in the absence of
the $b$-condensate, the one-particle Green functions for $b$
and $f$ fields are site-diagonal (or constant in the $\kk$-space)
to all orders in the perturbation
$(1/\sqrt{N})\sum_{\kk,\qq,\sigma}t(\kk) ( p_{\kk\sigma}\kr b_{\qq}\kr
f_{\kk+\qq\sigma}
+  f_{\kk+\qq\sigma}\kr b_{\qq}p_{\kk\sigma} ) $
in  \REF{HL}.
	In simple words, it is impossible to create (or annihilate)
a $b$ or $f$ particle at some site,
and then propagate it to some other site,
since that would violate the conservation of $Q_\nn$.

	Further following Kroha et al. \CITE{Kroha92} we confine
ourselves to a non-crossing approximation
(NCA) for which the free energy is
represented by diagrams  given in
Fig.~\FIGREF{Ffig}.
%\vbox to 0.5truecm{\vbox to 0.05cm{} Fig.~\FIGREF{Ffig}.}
	Fig.~\FIGREF{Gfig} shows the corresponding diagrams
for the full Green functions for $b$,
$f$ and $p$ fields.
	The set of Dyson equations reads as
\begin{eqnarray}
\label{Gb}
{\cal G}_b(\omega  )^{-1} & = &{\cal G}_b^0(\omega  )^{-1}-
{\it \Sigma}_b(\omega  ) \mbox{~,~~~~} \\
\label{Gf}
{\cal G}_f(\omega  )^{-1}& = & {\cal G}_f^0(\omega  )^{-1}-
{\it \Sigma_f}(\omega  ) \mbox{~,~~~~} \\
\label{Gc}
{\cal G}_p(\omega,\kk)^{-1}& = & {\cal G}_p^0(\omega  )^{-1}-
{\it \Sigma_p}(\omega,\kk) \mbox{~,~~~~}
\end{eqnarray}
with self energies ${\it \Sigma}_b$,
${ \Sigma}_p$ and ${\it \Sigma}_f$
of the following form
\begin{equl}{Sb}
{\it \Sigma}_b(\omega)={1\over N}
		\sum_\kk \mid t(\kk)\mid^2 \sigma_b(\omega,\kk)
\mbox{~,~~~~}
\sigma_b(\omega,\kk)\equiv{2\over \beta}
\sum_\nu {\cal G}_p(\nu,\kk) {\cal G}_f(\omega+\nu  )
\mbox{~,~~~~}
\end{equl}
\begin{equl}{Sc}
{\it \Sigma}_p(\omega,\kk)=\mid t(\kk)\mid^2 \sigma_p(\omega)
\mbox{~,~~~~}
\sigma_p(\omega)\equiv
{1\over \beta}\sum_\nu {\cal G}_b(\nu  ) {\cal G}_f(\omega+\nu  )
\mbox{~,~~~~}
\end{equl}
\begin{equl}{Sf}
{\it \Sigma}_f(\omega)={1\over N}
		\sum_\kk \mid t(\kk)\mid^2\sigma_f(\omega,\kk)
\mbox{~,~~~~}
\sigma_f(\omega,\kk)\equiv{1\over \beta}\sum_\nu {\cal G}_b(\nu  )
{\cal G}_p(\omega-\nu,\kk)
\mbox{~.~~~~}
\end{equl}
 The bare Green functions are
$$
{\cal G}_f^0(\omega)=\frac{1}{i\omega-\Lambda-\epsilon_d+\mu}
\mbox{~,~~~~}
{\cal G}_b^0(\omega)=\frac{1}{i\omega- \Lambda}
\mbox{~,~~~~}
{\cal G}_p^0(\omega)=\frac{1}{i\omega-\epsilon_p +\mu}
\mbox{~,~~~~}
$$
where $\mu$ denotes the chemical potential for holes.
%and $N$ stays for the number of unit cells in the crystal.
	The system of equations \REF{Gb}-\REF{Sf} has
to be solved
self-consistently  keeping at the same time
$\srv{Q_n}=1$.
	The number of holes per unit cell
is fixed by the doping level of the system, $n_p+n_d=1+\delta$.

	In solving the system of equations \REF{Gb}-\REF{Sf},
we transfer them, by means of usual methods of analytic
continuation \CITE{AGD} to the expressions in which the integrals
over various spectral functions appear.
	In the numerical integration over frequency a mesh of thousand
points per eV is used.
%
%	The procedure that we use in solving numerically the
%system of equations \REF{Gb}-\REF{Sf}
%starts by transferring the whole system of
%equations onto a discrete lattice in $\omega$.
%	Typically we use thousand points per eV.
	Starting with bare Green functions
we  iterate the complete set of equations until
the result does not change any more.
	Depending on the
iteration scheme, the rate of convergence may
change considerably, but the same
results emerge at the end.
	For numerical purposes, we use narrow Lorentzians
instead of the delta functions for the spectrum of the bare
Green functions for fermions.
	Their widths are typically several times of the mesh size.
	Once the Green function ${\cal G}_p(\omega,\kk)$ is known
we obtain the Green function for the real hole on the copper
site by using the relation
\begin{equl}{Gd}
{\cal G}_d(\omega,\kk)^{-1}=
{\it \sigma}_p(\omega)^{-1}-\mid t(\kk)\mid^2{\cal G}_p^0(\omega)
\mbox{~,~~~~}
\end{equl}
shown diagrammatically in Fig.\FIGREF{Gdfig}.
%\vbox to 0.5truecm{\vbox to 0.05cm{} Fig.~\FIGREF{Gdfig}.}

\section{Hole spectrum and the local density of states}

	In this section we present the results for
the  Green functions, which are obtained on solving
the system of equations \REF{Gb}-\REF{Gd}.
	For further discussion, we choose parameters \CITE{Mila88}
$\Delta_{pd}=\epsilon_p-\epsilon_d=4.0t_0$ ($t_0$
is of the order of 1eV),
$\delta=0.13$ extra holes per unit cell,
and set $\epsilon_p=0$ for convenience.
	The values for $\Lambda$
and $\mu$  that correspond  \CITE{mudelta} to
this choice of  parameters are
$\Lambda=3.75t_0$ and $\mu=-1.138t_0$.
	The chemical potential is positioned between the
energies $\epsilon_d$ and $\epsilon_p$  of the
copper and oxygen orbitals
and the temperature is set to $0.0135t_0$.
%The
%spectra ${\rm Im}
%G_d(\omega,\kk)$   and ${\rm Im}
%G_c(\omega,\kk)$
%hole density of states ($D.O.S.$) for the  copper and the
%oxygen site which we obtain for  $\delta=0.13$
%extra holes per unit cell  are shown in
%Fig.~\FIGREF{GdGcfig}.
%\vbox to 0.5truecm{\vbox to 0.05cm{} Fig.~\FIGREF{fbdosfig}.}

	For the $f$-fermion spectrum we get a simple
Lorentzian-like peak (Fig. \FIGREF{fbdosfig}),
with about 80\% of spectral weight in it,
positioned on the Fermi level and
with the occupancy $n_f=0.8$ (both spin projections).
	The structure of the $b$-boson spectrum is much
richer, with a high frequency peak centered at
%the renormalized
$\omega\sim\Lambda$,
%parameter (roughly corresponding to $\Delta_{pd}-t^2/\Delta_{pd}~{ }^*$),
%\HADD{* mozda malo neprecizno receno}
and with a tail extending to higher frequencies.
	For lower frequencies, we find
  a.) the asymmetric peak at $\omega\approx 0$ and
  b.) the broad plateau starting from $\omega\approx 0$
and extending beyond -1eV.
%\vbox to 0.5truecm{\vbox to 0.05cm{} Fig.~\FIGREF{dosGdGcfig}.}
	These features give rise to the following physical hole spectrum.

	The hole density of states per spin (hereafter DOS),
$-(1/N\pi)\sum_{\kk}{\rm Im}G^R(\kk,\omega)$, accumulates
(see Fig.~\FIGREF{dosGdGcfig}) around
energies  $\epsilon_p$ ({\it ``p-band''}),
$\epsilon_d$ ({\it ``d-band''}) and  $\mu$ ({\it ``in-gap-band''}).
While the accumulation near $\epsilon_p$ and
$\epsilon_d$ occurs for physically obvious reasons, the third region
may be related to the band that occurs around the
renormalized copper energy level in the
saddle point approximation \CITE{KLR88} for the $b$ field.
%	Physically, this corresponds to the {\it p-d} singlet
%formation, as described by Zhang and Rice \CITE{ZhangRice}.
	Experimentally, the ``in-gap'' states  show directly
as  a bump in the PES/IPES spectra \CITE{bumpEXP}
upon doping.
	 The dispersion that occurs in the ``band'' around
 $\epsilon_p$ is characterized by the shift of the
 peaks in ${\rm Im} G_p(\kk,\omega)$ towards higher hole
 energies along the $\Gamma X$ line in the Brillouin
 zone.
	This is in qualitative accordance with
 a simple tight-binding
 calculation for the $U_d=0$ case,
 except for the fact that now the peaks are
 of considerable width.
	The scattering of fermions on the slow boson field
resembles the scattering on the random potential and,
similarly to the crystal disorder, increases the
relaxation rate for the fermions.
	 The dispersions of the other two ``bands''
 go in the opposite direction (see Fig. \FIGREF{GdGcspfig}).
%\vbox to 0.5truecm{\vbox to 0.05cm{} Fig.~\FIGREF{GdGcspfig}.}
	Relative intensities
of DOS for the oxygen and copper site show that ``bands''
close to $\epsilon_p$ and $\mu$ are
dominantly of oxygen character, while
the ``band'' around $\epsilon_d$ is, as expected,
mainly of copper character.
 	The overall result is qualitatively similar
 to the DOS obtained by carrying out
 the Gaussian corrections beyond the saddle point approximation
 in the $1/N$ expansion for the generalized
 $N$-component model
\CITE{Melo}.
%\CITE{Dos1xN}.
%The contributions
%of different energy ranges to a local number
%of states may be also seen form the  functions
%$
%\int_{-\infty}^\omega  (d\nu/\pi)
% \sum_\kk \mid {\rm Im}\, G_{d,p}(\nu,\kk) \mid
%$
% which are also
%shown in  Fig.~\FIGREF{intGdGcspfig}.

	At this point the total number of states on
the copper and the oxygen atom can be found.
	The total number of states per spin
at the oxygen ``site'', given by the integral of
$-(1/\pi){\rm Im}G_c^R(\kk,\omega)$,
is 1, as it should be.
	However, the number of states per spin
per copper ``site'' should be 0.5,
as one state per  copper site is shifted to infinity
in  the $U_d\ide \infty$ limit, while the integral over
${\rm Im}G_d$ gives a somewhat larger number, $0.5\cdot [1+(1-n_f)]$.
	This deviation can be readily traced back to the
$\lambda_\nn=const$
approximation which we use and to the
decoupling \CITE{LuDecoupl}
of  $b$ and $f$ fields that takes place
in the calculation of  $\sigma_p$.
	Indeed, after the analytic continuation
to real frequencies, the equation \REF{Sc} gives
\begin{equl}{nfnb}
\begin{array}{ll}
  -\int {d\omega \over \pi} {\rm Im}\,
  \sigma^R_{p}(\omega)
& =-\int {d\omega \over 2\pi}
 \{{\rm Im}\, G^R_b(\omega  ) cth({\omega\over 2T}) -
    {\rm Im}\,G^R_f(\omega  )
    th({\omega\over 2T})   \} \\ [3mm]
  &  = {1\over 2}n_f+n_b = {1\over 2} + \frac{1-n_f}{2}
\mbox{~,~~~}
  \end{array}
\end{equl}
	These  extra $(1-n_f)/2$ states
%hybridize with oxygen states through
%Eqs. \REF{Gc} and \REF{Gd}
enter into the DOS through the
hybridization equations \REF{Gc} and \REF{Gd} producing
the aforementioned  deviation.

{  One feature of our results which seems alarming when
compared with the ARPES measurement of
the Fermi surface geometry  is that
the chemical potential $\mu$, as
obtained by the presented
selfconsistent calculation, is
positioned {\it above} the peaks in
${\rm Im} G(\kk,\omega)$. It means
that there is  practically  no Fermi
surface in our calculation, although
$\mu$ lays inside the in-gap band  since
the tails in ${\rm Im} G(\kk,\omega)$
extend above  $\mu$. The reason for
this may be traced back to the position of the peak in
${\rm Im}\,G_b(\omega)$ which is close to $\omega=0$.
The precise position of the peak determines the
Fermi surface geometry and the shift of this peak of the order of
0.05eV to lower frequency would cure the problem.
We believe that
the calculation beyond NCA  may produce such a shift
restoring the Fermi surface
consistent with the  experiment.
On the other hand,  at the energy scales larger than a
 tenth of an eV the major features
of the spectra would presumably remain unchanged
in the improved calculation. These energy scales are of the
prime interest in further discussion.
This concerns in particular the plateau in the
boson excitation
spectrum which extends over the range \CITE{OUReV}
of 1eV.  In the
calculation of the Raman spectra  this
plateau implies the occurrence of the
flat background. }

\section{Raman spectrum}

Before describing the calculation  of
the Raman spectrum we wish to recall
the simple results obtained   with the
electronic relaxation rate introduced
artificially \CITE{GfRam}.
Using the fermions with the Fermi liquid
like relaxation rate
$\gamma\approx\omega^2/\Gamma$
gives a Raman
spectrum with a {\it maximum} at the frequency of the order
$\Gamma$.
 Further on, such spectra with a maximum are obtained
for a wider class of the  non-Fermi liquid
electronic relaxation rate laws
$\gamma\approx \Gamma (\omega/\Gamma)^\alpha$
(with $\alpha>0$, $\alpha\neq 1$;
$\Gamma$ always being the characteristic
energy scale involved)
This is in contradiction with the experimentally observed
Raman scattering intensity which does not diminish
(or which even increases) towards high transfer
frequencies of 1eV.
   The phenomenological  marginal Fermi liquid {\it ansatz} of
Varma et al.  \CITE{VarmaMFL89} ($\alpha= 1$ in the above
$\gamma$)  has no   such parameter $\Gamma$ and a linear
damping rate for electrons leads to the constant
Raman background.

{ In what follows we do not use any artificial
assumption for the electronic spectrum. Instead, we
use the boson and the fermion propagators obtained
by the numerical calculation of the preceding section.
While it is difficult to draw the correspondence
between the aforementioned phenomenological
calculation and the ours,
%electronic
%spectra and those of our calculation
%(partly because of the problems that we
%encounter with the electronic spectrum within the
%$0.1eV$ from the Fermi level)
it is interesting to note that our plateau in the slave
boson spectral function is the feature
which leads to the nearly constant Raman background
similar to that  obtained in
the marginal Fermi liquid phenomenology
\CITE{VarmaMFL89}.}

{ The calculation of the Raman spectrum proceeds
as follows.}
%
%	  It may be noted here that the
% electronic spectra that  we obtain above
% show a significant damping in the
% ``in-gap-band'' (ie. close to the Fermi level).
% 	  However, that damping
% is too small to be  responsible for a
% Raman background up to $1eV$.
% 	 It may be further  thought  that   corrections  of
% our results beyond the NCA and the inclusion
% of  fluctuations of the $\lambda$ field
% may act to reduce this damping and make
% the whole electronic picture  more
% ``Fermi liquid like''.
% 	 While these considerations are
% beyond the scope of the present paper, it should be
% noted that this would, presumably, not affect the Raman spectrum
% at higher frequencies, for which the present
% calculation accounts very well, as shown below.
%\def\baselinestretch{2.0}
%\baselineskip=16pt
	The coupling of  the electromagnetic field
to electrons is introduced, as usual, by the substitution
\begin{equl}{addA}
t_{r,s}(c_r^{\dagger}c_s+c_s^{\dagger}c_r)
\ide t(c_r^{\dagger}c_se^{\FAZA{s}{r} }
+c_s^{\dagger}c_re^{\FAZA{r}{s} })
\mbox{~,~~~}
\end{equl}
in the interatomic hybridization term of the original
Hamiltonian.
	This is followed by the expansion of the exponentials
up to the terms quadratic in the vector potential $\vec{A}$.
	The terms $\vec{p}\cdot \vec{A}$ and  $n{\vec{A}}^2$
describe  the coupling of the electromagnetic field to the
hole momentum and to the bond charge.
	As usual, these terms
give rise to three different contributions $T_a$, $T_b$ and $T_c$
to the transition matrix for the Raman scattering.
	They are shown in Fig.~\FIGREF{Tmatfig} and correspond to
%\vbox to 0.5truecm{\vbox to 0.05cm{} Fig.~\FIGREF{Tmatfig}.}
\begin{itemize}
\item[a)] $n{\vec{A}}^2$ scattering;
\item[b)] $(\vec{p}\cdot \vec{A})^2$ scattering where
electron first absorbs the incoming photon with  frequency
$\omega_I$ and then emits the outcoming photon with frequency
$\omega_S$;
\item[c)] similar to b) but with absorption and
emission processes  taking place in the reverse time order.
\end{itemize}

	The  probability for the transition
between the initial state
$\braket{0}$ (for simplicity we use zero temperature
formulae to illustrate further steps) and
the final states $\braket{f}$ consistent with
the Raman processes described above
is equal to
\begin{equl}{Wt}
  W={2 \pi \over \hbar}\sum_f
  {\mid (T_a+T_b+T_c)_{0f} \mid^2\delta(E_0-E_f) }
\mbox{~.~~~}
\end{equl}
 This gives six (three diagonal and three nondiagonal)
 contributions to the transition probability,
\begin{equl}{Wall}
  W=W_{aa}+W_{bb}+W_{cc}+W_{ab}+W_{ac}+W_{bc}
\mbox{~,~~~}
\end{equl}
%\begin{equl}{Wdia}
%  W_{ii}={2 \pi \over \hbar}\sum_f{T_i^*T_i\delta(E_0-E_f) }
%\mbox{~,~~~}
%\end{equl}
%\begin{equl}{Wndia}
%  W_{ij}(i\neq j)={2\pi \over \hbar}
%                \sum_f{(T_i^*T_j+T_j^*T_i)}\delta(E_0-E_f)
%\mbox{~,~~~}
\begin{equl}{Wndia}
  W_{ij}=(2-\delta_{ij}){\pi \over \hbar}
                \sum_f{(T_i^*T_j+T_j^*T_i)}\delta(E_0-E_f)
\mbox{~,~~~}
i,j=a,b,c
\mbox{~.~~~}
\end{equl}
%
%The expressions for $W_{ij}$ may be
%pictured diagrammatically (e.g.. see Fig.~\FIGREF{Taafig} for
%$W_{aa}$) and written down in a
%straightforward manner (more detailed expressions for the
%electron-light coupling are  supplied in Appendix C).
%\mxnote{C?-to be changed}

	The Raman spectra that emerge after the
numerical integration are shown in Fig.~\FIGREF{Rampolfig}.
	The featureless behaviour that
characterizes the resulting Raman spectrum
extends approximately up to the frequency
of the separation of the ``in-gap-band''
 and the  ``p-band''.
	This separation is of the order of 1eV.
	This agrees with experiment, while
scatterings for larger frequencies are not
experimentally probed.
	Above  that frequency, our calculations give
 a sudden increase of the Raman intensity,
due to the ``interband'' processes.
%\vbox to 0.5truecm{\vbox to 0.05cm{} Fig.~\FIGREF{Rampolfig}.}
	Coming back to the spectrum below 1eV,
we note that the intensity of the Raman spectrum
changes significantly as a function of
the polarization of  the incident and the
scattered light.
	This is  due to the
polarization dependence of the
electron-light  interaction  which comes from
the nearest-neighbour hopping approximation
(see the Appendix for some details).
%	The spectrum for some combinations
%of polarizations  that typically occur in
%the experiment \CITE{SHM1,SHM2}
%(with wave vector of the
%incoming and the scattered light
% perpendicular to the
%copper-oxide planes) are shown in Fig.~\FIGREF{Ramexpfig}.
	An example of the relative size of  contributions which come
from various terms in  \REF{Wall} is shown in
Fig.~\FIGREF{Ramsizefig}.
%\vbox to 0.5truecm{\vbox to 0.05cm{} Fig.~\FIGREF{Ramsizefig}.}
	The relative importance of various terms
changes with polarization
(however, the diagonal terms almost always dominate over
the nondiagonal ones).
	In particular, the scattering via the
$n{\vec{A}}^2$ coupling does not
contribute at all for the polarization combination
 $(x',y')$.
	In general, the polarization dependence
of the  intensity of the Raman spectrum
shows  good qualitative agreement with
experimental results.
	The minor
difference in that respect is that the
intensity of our spectra for  $(x,x)$ and
$(y',y')$
polarizations  are
somewhat lower that the experimental ones \CITE{SHM2}.
	This may point to the necessity of including
direct oxygen-oxygen hybridization terms
(as well as the corresponding electron-light coupling terms)
in the initial {\it p-d} model.

\vbox to 0.5truecm{}

	In summary, we have calculated the hole spectrum in the
CuO$_2$ plane, using the slave boson formalism to
implement the constraint of no double occupancy on the copper site,
and without resorting to the usual saddle point
approximation for the slave boson field.
%	For the copper-oxide superconductors,
%the Green functions for
%the holes in CuO$_2$ planes and Raman spectra
%are calculated.
%	Because of large copper on-site Coulomb
%repulsion ($U_d$) slave boson formalism is applied.
	A three band density of states structure is obtained, with
the Fermi level lying in the ``in-gap'' band.
	The corresponding energy scales show up in the Raman spectra,
featureless in frequency up to the value of the gap between
the ``in-gap-band'' and the ``p-band''.
	The measured polarization dependence agrees well
with our calculation, which is based on the assumption that the Cu-O
hopping $t_0$ is the dominant one in the CuO$_2$ planes.
%
%	The hole Green functions were further used
%to calculate the Raman spectra, featureless
%in frequency and with specific polarization dependence,
%both in qualitative agreement with experiments.
%

\vskip 10mm

{ This work was supported by the Croatian Government under
project 1-03-275 and by the EU-Croatia
project  CI1*0568-C(EDB). }
\appendix

\section{Appendix}

%\def\baselinestretch{2.0}
%\baselineskip=16pt

% Here we give list of polarization dependencies
%for nonzero vertexes for the Raman scattering.
% With nearest-neighbour hopping \REF{H_hyb_cd}
%hopping as we have chosen only the following transitions are
%nonzero for the polarization.
% P letter means
% polarization of incoming ($I$) or outcoming ($S$)
% electromagnetic wave.
% Letters $d$, $p$, $c$ are shorthange for copper-site hole,
% oxygen-site hole (bonding combination of the hole operators)
% and oxygen-site hole with nonbonding combination
% of the operators respectively.

	The polarization dependence
(and the corresponding symmetry properties)
of the Raman spectrum result directly from
the polarization dependence of the electron-light
coupling terms in the $\vec{k}$-space representation.
	The coupling constants $g_1$ and $g_2$ correspond
respectively to $\vec{p}\cdot\vec{A}$
and $n\vec{A}^2$ terms.
In the {\it p-d} model in which only Cu-O
hopping is included these coupling
constants have the following form:
%	With electron-light interaction including
%~$\vec{p}\cdot\vec{A}$~ and ~$n\vec{A}^2$~ terms,
%the coupling constants describing
%hole-momentum-light coupling ($g_1$)
%and bond-charge coupling ($g_2$) emerge.
%	Thereby, copper-site ($d$) $\leftrightarrow$
%oxygen-site ($p$ for bonding and $n$ for nonbonding orbital)
%transitions occur.
%	With $P$ denoting polarization index
%($\hat{e}_p$ stands for polarization vector),
%and $I$ for incident and $S$ for scattered light,
%coupling constants are the following:
%
\begin{equation}
 g_{1,\kk}^{dn}
%=-
\propto-
%=-{e^2\over 8\hbar c^2}
%{1\over \sqrt{\mid t_{x\kk}\mid^2+\mid t_{y\kk}\mid^2}
{[(\hat{e}_p\cdot \hat{x})t_{y\kk}\kappa_{x\kk}^*
-(\hat{e}_p\cdot \hat{y})t_{x\kk}\kappa_{y\kk}^*]
\over \sqrt{\mid t_{x\kk}\mid^2+\mid t_{y\kk}\mid^2}}
%~,~P={\rm polarization}
\mbox{~,~~~}
\end{equation}
\begin{equation}
 g_{1,\kk}^{dp}
%=-
\propto-
%=-{e^2\over 8\hbar c^2}
%{1\over \sqrt{\mid t_{x\kk}\mid^2+\mid t_{y\kk}\mid^2}
{[(\hat{e}_p\cdot \hat{x})t_{x\kk}\kappa_{x\kk}^*
 +(\hat{e}_p\cdot \hat{y})t_{y\kk}\kappa_{y\kk}^*]
\over \sqrt{\mid t_{x\kk}\mid^2+\mid t_{y\kk}\mid^2}}
%~,~P={\rm polarization}
\mbox{~,~~~}
\end{equation}
\begin{equation}
 g_{2,\kk}^{dn}
%=-
\propto-
%{e^2\over 8\hbar c^2}
%{1\over \sqrt{\mid t_{x\kk}\mid^2+\mid t_{y\kk}\mid^2}
{[ (\hat{e}_I\cdot \hat{x})(\hat{e}_S\cdot \hat{x})
 -(\hat{e}_I\cdot \hat{y})(\hat{e}_S\cdot \hat{y})]t_{x\kk}t_{y\kk}
\over \sqrt{\mid t_{x\kk}\mid^2+\mid t_{y\kk}\mid^2}}
\mbox{~,~~~}
\end{equation}
\begin{equation}
 g_{2,\kk}^{dp}
%=-
\propto-
%{e^2\over 8\hbar c^2}
%{1\over \sqrt{\mid t_{x\kk}\mid^2+\mid t_{y\kk}\mid^2}
{[(\hat{e}_I\cdot \hat{x})(\hat{e}_S\cdot \hat{x})\mid t_{x\kk}\mid^2
+(\hat{e}_I\cdot \hat{y})(\hat{e}_I\cdot \hat{y})\mid t_{y\kk}\mid^2]
\over \sqrt{\mid t_{x\kk}\mid^2+\mid t_{y\kk}\mid^2}}
\mbox{~,~~~}
\end{equation}
where
\begin{equation}
\kappa_{x\kk}\equiv it_0(1+e^{-ik_x}) \mbox{~,~~~}
\kappa_{y\kk}\equiv it_0(1+e^{-ik_y})~
\mbox{~.~~~}
\end{equation}
	Here $\vec{e_I}$ and $\vec{e_S}$ denote polarization
vectors of the incident and scattered light
(in $g_1$ expressions $\vec{e_P}$=$\vec{e_I}$ or $\vec{e_S}$).
The superscripts {\it dp} and {\it dn} stand
for the case of the coupling of the light to the copper-bonding
and the copper-nonbonding oxygen terms respectively.

\newpage

\pagestyle{empty}

\def\RCITE#1{\bibitem{#1}}

\vfill \eject

\begin{itemize}

\item[Fig.~1:~]First few diagrams for the free energy within the $NCA$.

\bigskip

\item[Fig.~2:~]The diagrams for the Dyson equations for $b$, $f$ and $p$
fields within the NCA approximation. $f$ and $b$ are site-diagonal. Bold
lines denote the full Green functions.

\bigskip

\item[Fig.~3:~]The diagram for the Green function for the hole on the copper
site.

\bigskip

\item[Fig.~4:~]The spectra for the (a) $f$-fermion and (b) $b$-boson.

\bigskip

\item[Fig.~5:~]DOS for (a) the oxygen site and (b) the copper site holes.

\bigskip

\item[Fig.~6:~]In-gap band dispersion (the one containing the Fermi level)
along the $\Gamma X$ line. The spectra for (a) the oxygen site and (b)
the copper site are shown.

\bigskip

\item[Fig.~7:~]Diagrams for the lowest order contributions to the T-matrix
for the Raman scattering.

\bigskip

\item[Fig.~8:~]Transition probability for the Raman scattering for various
polarizations of the incident and scattered light as the function
of the transfer energy $\omega$.

\bigskip

\item[Fig.~9:~]The contributions ($W_{aa}$, $W_{bb}$, $W_{cc}$ in (a) and
$W_{ab}$, $W_{ac}$ $W_{bc}$ in (b)), coming from various T-matrix terms,
to the total transition probability $W$ are shown for the $(x,x)$
polarization.

\end{itemize}

\vfill

\eject

 \vbox to 4.0truecm{}
%PRVA SLIKA NEKOLIKO DIJAGRAMA ZA SLOBODNU ENERGIJU

               \begin{figure*}[htb]

\centerline{\psfig{figure=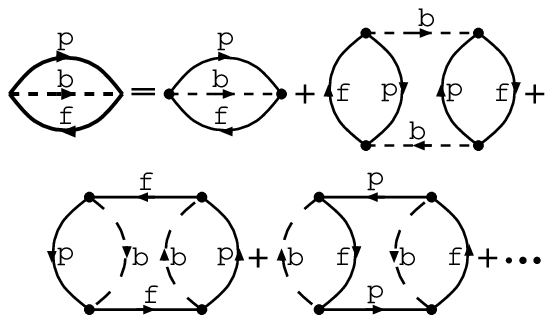,width=7.6cm,height=4.2cm}}
 \vbox to 4.0truecm{}
               \caption{~ \label{Ffig}}
               \end{figure*}

 \vfill

\eject

 \vbox to 4.0truecm{}
%DRUGA SLIKA NEKOLIKO DIJAGRAMA ZA SLOBODNU ENERGIJU

               \begin{figure*}[htb]

\centerline{\psfig{figure=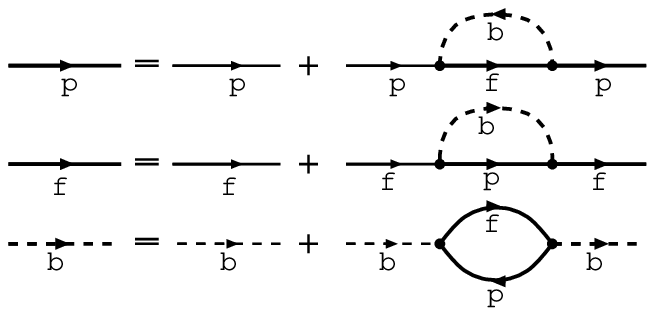,width=7.6cm,height=3.8cm}}
 \vbox to 4.0truecm{}
               \caption{~ \label{Gfig}}
               \end{figure*}

 \vfill

 \eject

 \vbox to 4.0truecm{}
%TRECA SLIKA NEKOLIKO DIJAGRAMA ZA SLOBODNU ENERGIJU

               \begin{figure*}[htb]

\centerline{\psfig{figure=clan_hn1_f_Green2.eps,width=7.6cm,height=1.3cm}}
 \vbox to 4.0truecm{}
               \caption{~ \label{Gdfig}}
               \end{figure*}

 \vfill

 \eject

 \vbox to 4.0truecm{}
%CETVRTA SLIKA NEKOLIKO DIJAGRAMA ZA SLOBODNU ENERGIJU

               \begin{figure*}[htb]

\centerline{\psfig{figure=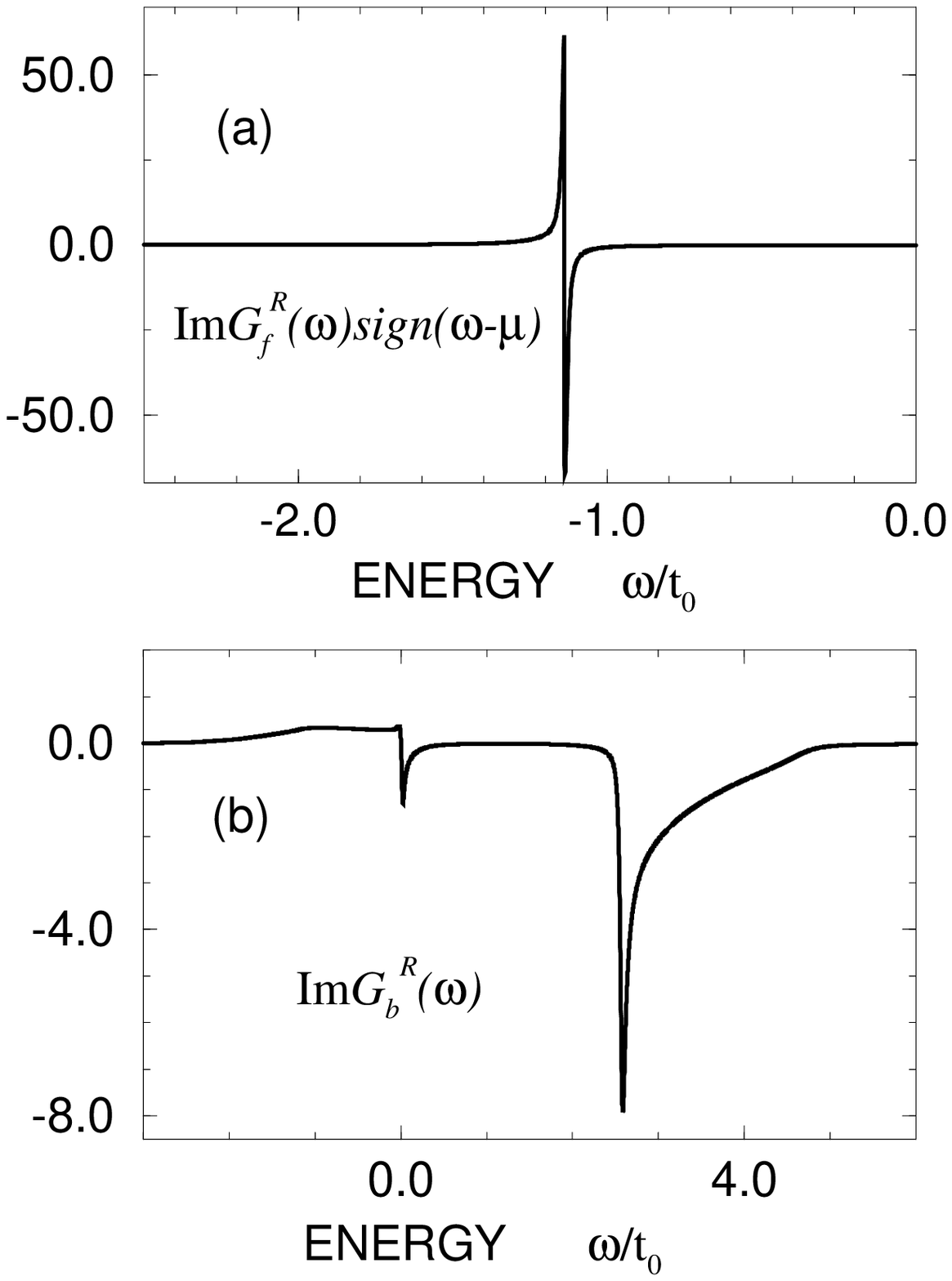,width=11.5cm,height=10.3cm}}
 \vbox to 4.0truecm{}
               \caption{~ \label{fbdosfig}}
               \end{figure*}

 \vfill

 \eject

 \vbox to 4.0truecm{}
%PETA SLIKA NEKOLIKO DIJAGRAMA ZA SLOBODNU ENERGIJU

               \begin{figure*}[htb]

\centerline{\psfig{figure=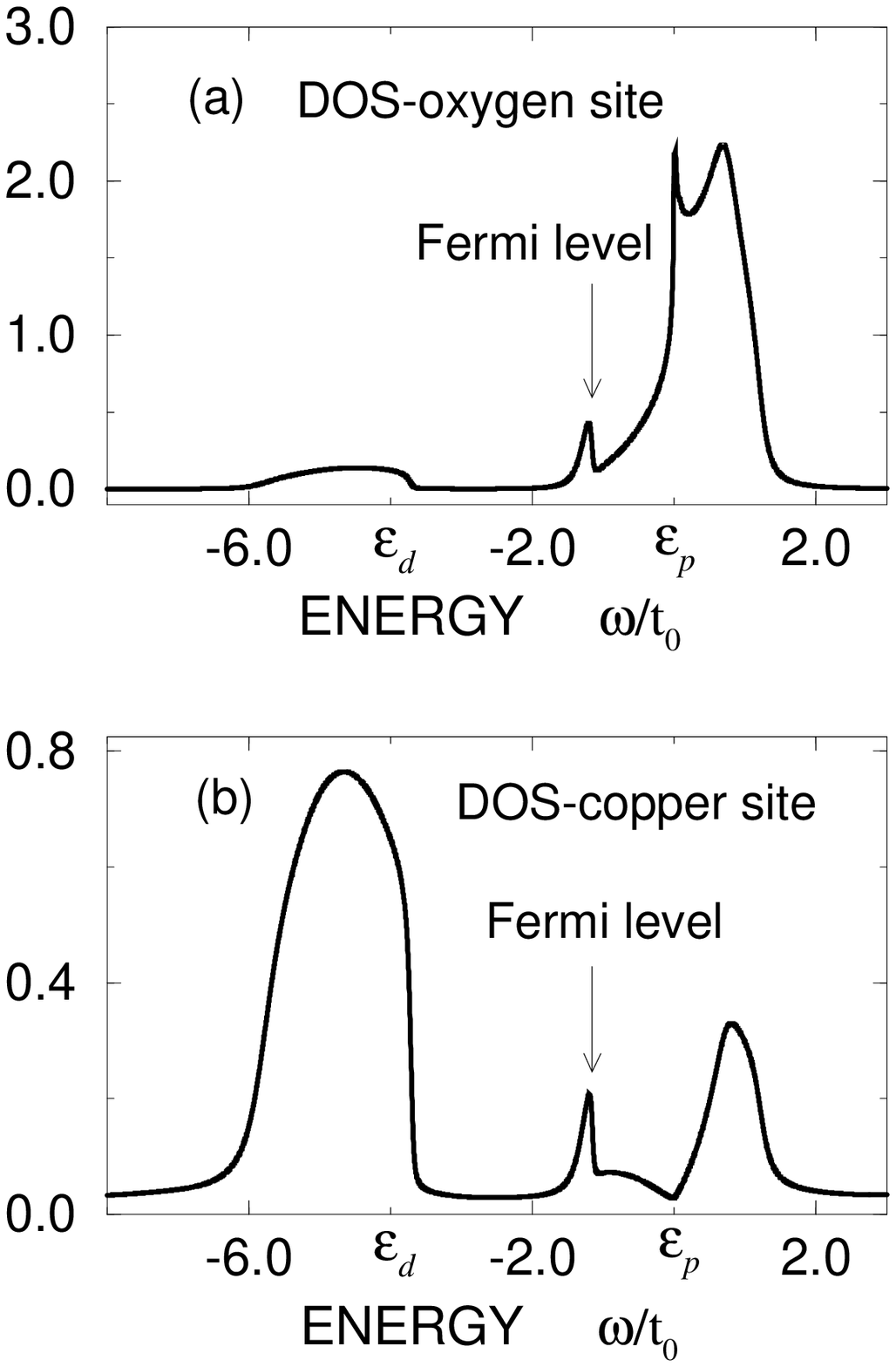,width=11.5cm,height=11cm}}
 \vbox to 4.0truecm{}
               \caption{~ \label{dosGdGcfig}}
               \end{figure*}

 \vfill

 \eject

 \vbox to 4.0truecm{}
%SESTA SLIKA NEKOLIKO DIJAGRAMA ZA SLOBODNU ENERGIJU

               \begin{figure*}[htb]

\centerline{\psfig{figure=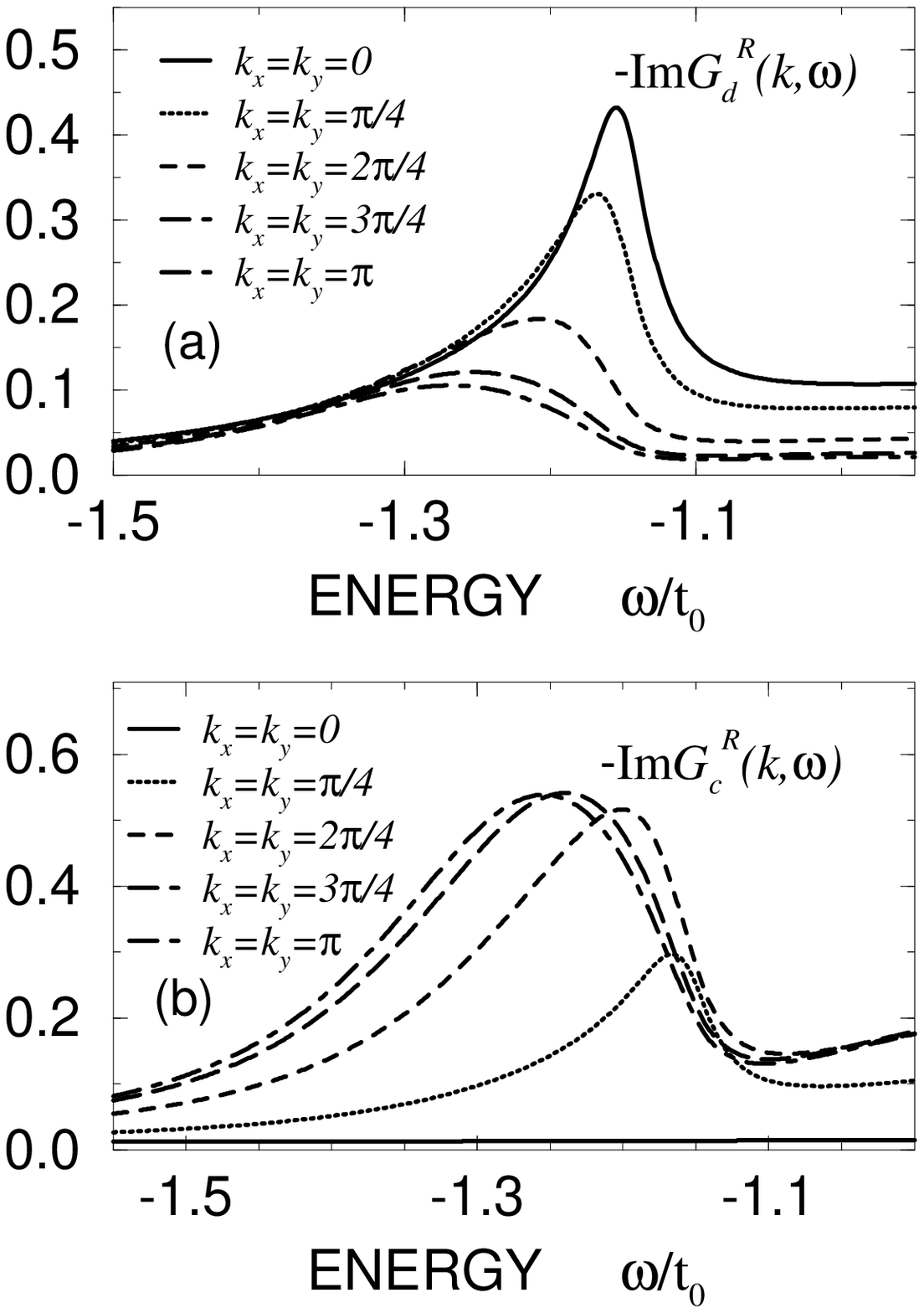,width=12.5cm,height=11.0cm}}
 \vbox to 4.0truecm{}
               \caption{~ \label{GdGcspfig}}
               \end{figure*}

 \vfill

 \eject

 \vbox to 4.0truecm{}
%SEDMA SLIKA NEKOLIKO DIJAGRAMA ZA SLOBODNU ENERGIJU

               \begin{figure*}[htb]

\centerline{\psfig{figure=clan_hn1_f_Tmat.eps,width=7.6cm,height=3.8cm}}
 \vbox to 4.0truecm{}
               \caption{~ \label{Tmatfig}}
               \end{figure*}

 \vfill

 \eject

 \vbox to 4.0truecm{}
%OSMA SLIKA NEKOLIKO DIJAGRAMA ZA SLOBODNU ENERGIJU

               \begin{figure*}[htb]

\centerline{\psfig{figure=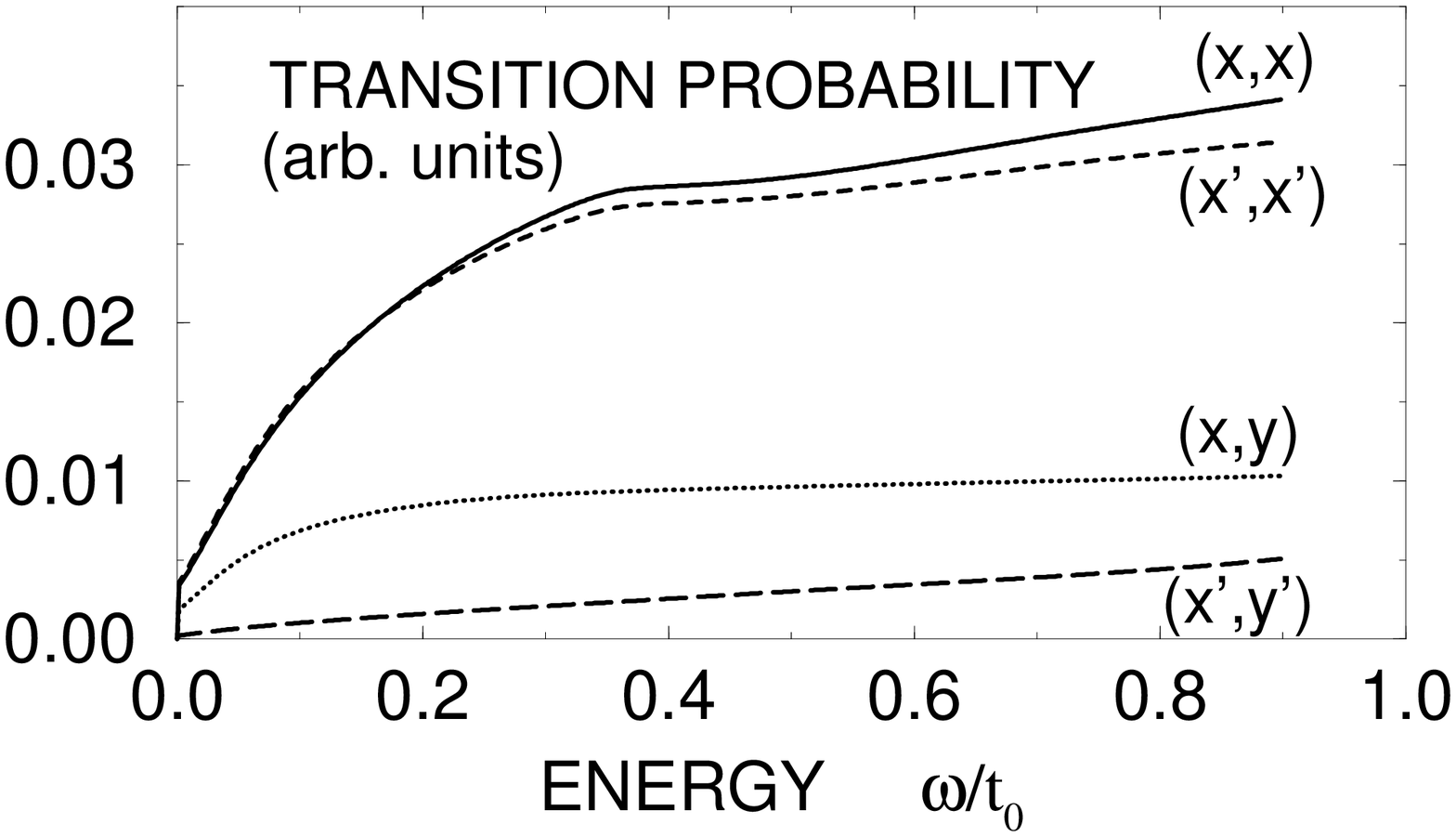,width=7cm,height=5.6cm}}
 \vbox to 4.0truecm{}
               \caption{~ \label{Rampolfig}}
               \end{figure*}

 \vfill

 \eject

 \vbox to 4.0truecm{}
%DEVETA SLIKA NEKOLIKO DIJAGRAMA ZA SLOBODNU ENERGIJU

               \begin{figure*}[htb]

\centerline{\psfig{figure=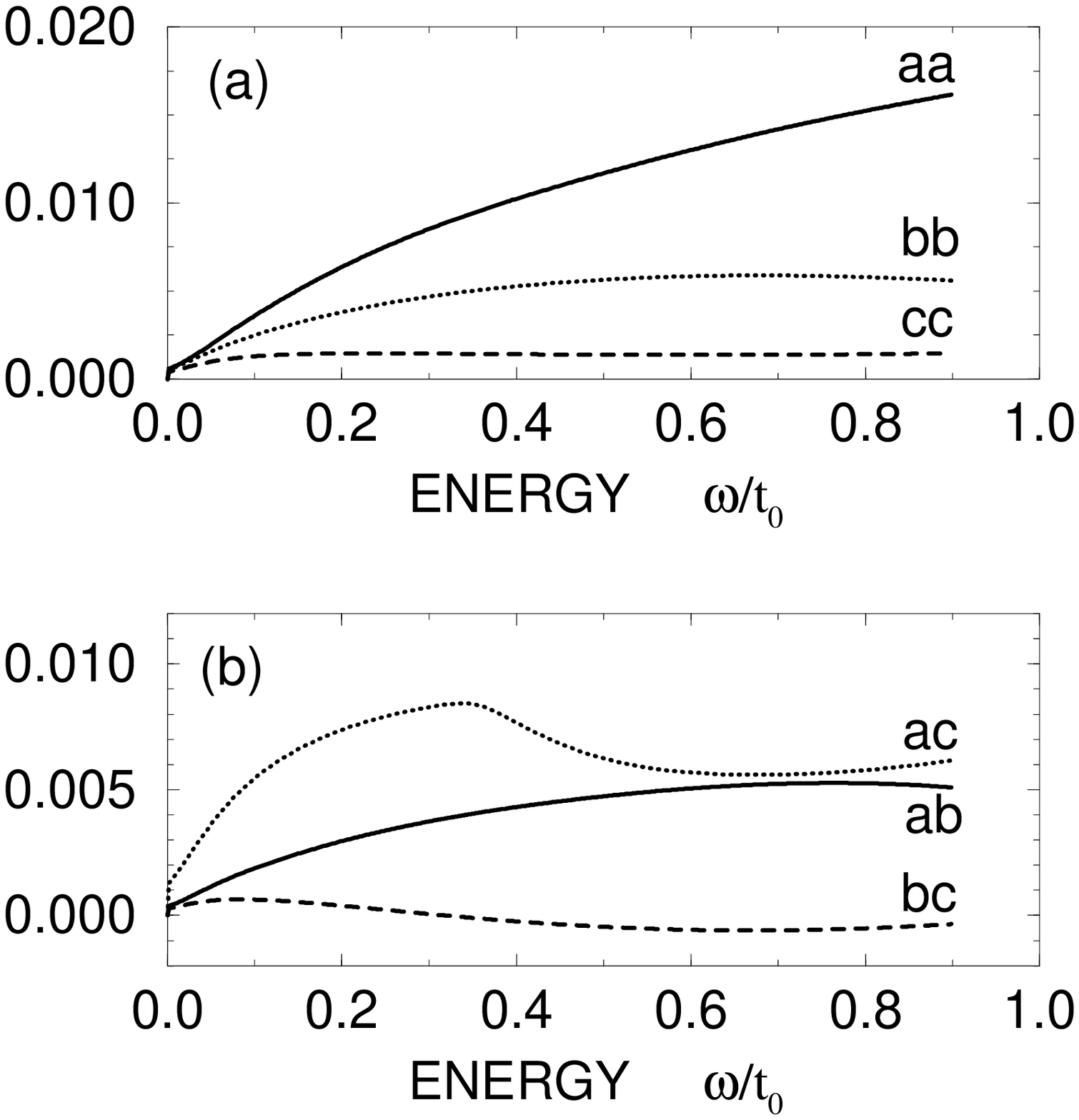,width=8.5cm,height=7.0cm}}
 \vbox to 4.0truecm{}
               \caption{~ \label{Ramsizefig}}
               \end{figure*}

 \vfill

 \eject

\end{document}